\documentstyle[12pt]{article}
\begin{document}
\renewcommand{\theequation}{\thesection.\arabic{equation}}

\vspace*{1cm}
\begin{center}
{\LARGE\bf Effective action for high-energy scattering}
\end{center}
\begin{center}

{\LARGE\bf in gravity \footnote{Work supported in part by the
Volkswagen Stiftung}}
\end{center}
\vspace{2.0cm}
\begin{center}
{\large\bf R. Kirschner}\\{\large\it International School for
Advanced Studies (SISSA / ISAS)}\\{\large\it via Beirut 2-4,
I-34013 Trieste,
Italy}
\end{center}
\vspace*{1.0cm}
\begin{center}
{\large\bf L. Szymanowski}\\{ \large\it Soltan Institute for Nuclear
Studies }\\{\large\it Hoza 69, PL-00681 Warsaw, Poland}
\end{center}
\vspace*{2.0cm}
\begin{center}
{\bf Abstract }
\end{center}
The multi-Regge effective action is derived directly from the
linearized gravity action. After excluding the redundant field
components we separate the fields into momentum modes and integrate
over modes which correspond neither to the kinematics of scattering
nor to the one of exchanged particles. The effective vertices of
scattering and of particle production are obtained as sums of the
contributions from the triple and quartic interaction terms and
the fields in the effective action are defined in terms of the
two physical components of the metric fluctuation.

\vspace*{1.0cm}
\begin{center}
SISSA Ref. 194/94/EP  (Dec. 94)
\end{center}

\newpage
\section{Introduction}

The study of high-energy scattering in gravity is considered as
a way to learn
about the yet unknown quantum theory of gravitation. At energies
of the Planck
scale quantum gravity effects become important.

At high energy and small momentum transfer the elastic scattering
is described
by the eikonal approximation. In this approximation the amplitude
can be
obtained by summing all graphs with the exchange of an arbitrary
number of non-interacting gravitons between the scattering particles
\cite{ACV87}, \cite{MS} as well as from the classical gravitational
shock wave solution \cite{tH}, i.e. the gravitational field of a
particle moving with the speed of light \cite{AiS}, \cite{DtH}.

In Yang-Mills theories the contributions of s-channel multi-particle
intermediate states dominate the contributions of eikonal type.
Unlike this in gravity, simply because of the higher spin, the
exchange of one more graviton results in an additional power of
$s$ and the contribution of multi-particle intermediate states
appears as a correction to the eikonal approximation.
Because the eikonal contributions sum up to a phase, these
corrections are more important than it seems from the first glance
to the perturbative expansion. Quantum effects enter in fact just
with these multi-particle contributions.

Corrections to the eikonal have been calculated \cite{ACV90},
\cite{ACV93}, \cite{FPVV}. There is an approach to the improved
eikonal \cite{ACV93}, where the multi-Regge effective action
is used. This action involves the effective vertices of scattering
and particle production appearing in the multi-particle
amplitudes at high energy with all pairs of particles in $s$-channel
having large sub-energies, i.e. in the multi-Regge kinematics.

The multi-particle amplitudes in this kinematics can be
obtained from the elastic amplitude at high energy by $t$-channel
unitarity. It is enough to know the elastic amplitude to get by
unitarity and gauge invariance the effective vertices \cite{BFKL},
\cite{FS}, \cite{L82} and the multi-Regge effective action
\cite{L91}. The effective vertices can also be obtained from
string amplitudes \cite{L88}, \cite{ABC}.

In the case of Yang-Mills theory (including fermions, QCD)
Lipatov and the authors have found a way to derive the
multi-Regge effective action  directly from the original
action \cite{KLS}. We write the action in the axial gauge,
choosing the momentum of one of the incoming particles as
the gauge vector.  After eliminating the redundant fields
we split the fields into parts, corresponding to momentum ranges
determined by the multi-Regge kinematics. The essential step
is the (approximate) integration over the "heavy" modes.

In the case of gravity the direct relation in this spirit between
the original action and the multi-Regge effective action has not
been investigated. The aim of the present paper is to fill
this gap.

The multi-Regge effective action is a tool to study the
high-energy peripheral scattering both in gravity and in
Yang-Mills theory. In the latter case it allows to reproduce
easily the results of the leading logarithmic approximation
(gluon reggeization, perturbative pomeron) and provides the
basis for a systematic improvement (generalized leading
logarithmic approximation) including the exchange of an
arbitrary number of reggeized gluons interacting with each
other in order to obey the of unitarity in all sub-energy
channels.

There is an effective action for high-energy peripheral
scattering both in Yang-Mills theory and in gravity
\cite{VV} obtained by shrinking the longitudinal dimensions.
It reproduces the eikonal approximation and the first
correction involving the effective particle production
vertex. The contributions with more than one
additional particle in the $s$-channel intermediate state
deviate from the ones from the multi-Regge effective
action. In particular the leading logarithmic approximation
is not reproduced.

In the present paper we extend our procedure of separating modes
and of integrating over "heavy" modes to the case of pure gravity.
We choose the axial gauge with the momentum of an incoming
particle as the gauge vector. The physical degrees of freedom
can be represented by two independent matrix elements
$\gamma_{11}, \gamma_{12} $, where $\gamma_{ij}$ is defined by the
transverse components of the metric $g_{ij}, i,j = 1,2 $
\begin{equation}
g_{ij} = e^{\psi } \ \ \gamma_{ij},   \ \ \ \ \ \
\det (\gamma_{ij} ) = 1.
\end{equation}
The elimination of the redundant field components in gravity
is much more involved compared to the Yang-Mills theory. The
result appropriate for our purposes is obtained \cite{SS},
\cite{Kaku}, \cite{AChF} by specifying the gauge fixing as
follows,
\begin{equation}
g_{--} = g_{-i} = 0, \ \ \ g_{-+} = e^{\psi /2 } .
\end{equation}
This leads to constraints, which can be solved in closed form
to eliminate $g_{++}, g_{+i}$ and $\psi $. In particular one finds
\begin{equation}
\psi = \frac{1}{4} \partial_{-}^{-2}
[\partial_{-} \gamma^{ij} \partial_{-} \gamma_{ij} ].
\end{equation}
The action is determined by the following Lagrangian \cite{SS},
\begin{eqnarray} 
16 \pi G \ {\cal L} =
e^{\psi} (4 \partial_+ \partial_- \psi - \partial_+ \gamma^{ij}
\partial_- \gamma_{ij} )   \nonumber \\
- e^{\psi /2} \gamma^{ij}
(\frac{1}{2} \partial_i \partial_j \psi - \frac{3}{8}
\partial_i \psi \partial_j \psi
- \frac{1}{4} \partial_i \gamma^{kl} \partial_j \gamma_{kl}
+ \frac{1}{2} \partial_i \gamma^{kl} \partial_k \gamma_{jl} )
                           \nonumber  \\
- \frac{1}{2} e^{-3/2 \psi} \gamma^{ij} \partial_-^{-1} R_i
\partial_-^{-1} R_j ,
                           \nonumber \\
R_i = \frac{1}{2} e^{\psi }
(\partial_- \gamma^{jk} \partial_i \gamma_{jk} +
\partial_i \psi \partial_- \psi -
3 \partial_- \partial_i \psi ) +
\partial_k (e^{\psi } \gamma^{jk} \partial_- \gamma_{ij} ).
\end{eqnarray}
We parametrize
\begin{equation}
\gamma_{ij} = ( e^h )_{ij}, \ \ \ \ \ \ {\rm Sp} \ h = 0,
\end{equation}
and use the complex field defined by the two independent elements
of the matrix $h$ by
\begin{equation}
h = {1 \over \sqrt 2 } (h_{11} +  i h_{12} ).
\end{equation}
Complex notations will be used also for two-dimensional transverse
momentum and position vectors as in \cite{KLS}. Our notations are
close to the ones in \cite{BBC}. The representation of the linearized
action given in the latter paper turned out to be a good starting
point for our analysis.
\begin{eqnarray} 
&{\cal L} = {\cal L}^{(2)} + {\cal L}^{(3)} + {\cal L}^{(4)} + ...,
 \nonumber  \\
&{\cal L}^{(2)} = - 2 h^* (\partial_+ \partial_-
-\partial \partial^* ) h,
 \nonumber  \\
&{\cal L}^{(3)} =  2 \alpha \{
(\partial_- h^* \partial_- h ) \partial^{* 2} \partial_-^{-2} h +
\partial_- h^* h \partial^{* 2} \partial_-^{-1} h -
2 \partial_- h^* \partial^* h \partial^{*} \partial_-^{-1} h
+ {\rm c.c.} \},
  \nonumber  \\
&{\cal L}^{(4)} = 2 \alpha^2 \{
- 2 \vert \partial_-^{-2}
( \partial_-^3 h^* \partial^* \partial_-^{-1} h -
\partial_-^2 \partial^* h^* h )  \vert^2 +
\vert  \partial_-^{-2} ( \partial_-^2 h^* \partial^* h -
\partial_- \partial^* h^* \partial h) \vert^2     \nonumber \\
& + \vert \partial_-^{-1} (\partial_- h^* \partial^* h -
\partial^* h^* \partial_- h) \vert^2
- 3 \vert \partial_-^{-1} (\partial_- h^* \partial^* h ) \vert^2
 \cr
 &+ 3 \partial_-^{-1} (\partial_- h^* \partial_- h)
\partial_-^{-1}(\partial h^* \partial^* h)
 \cr
 & + [ {\partial }_-^{-2} ( \partial_- h^* \partial_- h ) - h^* h ]
[ \partial h^* \partial^* h + \partial^* h^* \partial h -
\partial \partial^* \partial_{-}^{-1} h^* \partial_- h -
\partial_- h^* \partial \partial^* \partial_{-}^{-1} h ] \}.
\end{eqnarray}
A factor $(8 \pi G)^{1/2}$ has been included into $h$, $
\alpha = (4 \pi G )^{1/2}$.

We understand the inverse derivatives as operators defined by Fourier
transformation from the momentum representation with the zero mode
excluded.

Compared to the Yang-Mills case our analysis here is more involved
not only because there are more interaction terms but also
because the integration over the "heavy" modes has to be performed
with higher accuracy.

After introducing the separation of modes we study first the
contributions in
(1.7), which are relevant for the elastic and
inelastic scattering of a high-energy particle
in an external gravitational field. The source of this field is
actually the other incoming particle. It is natural to choose the
momentum of that particle as the gauge vector. It
can be useful to think of such processes, when reading sections
2, 3 and 4. We obtain the effective vertices for scattering (for
particles with momenta close to the momentum of the
incoming particle which is not the
gauge vector) and for particle production. For the latter vertex
we have first to derive the induced vertex arising from the
integration over the "heavy" modes.

In sections 5 and 6 we study the contributions in (1.7), which
are necessary beyond the ones considered before for elastic
and inelastic particle-particle scattering. The resulting
effective vertex for scattering with momenta close to the one
of the other incoming particle, the momentum of which is
the gauge vector, can be written without derivation by parity
symmetry. To obtain this vertex as a sum of contributions
from the quartic terms ${\cal L}^{(4)}$, the original triple
vertices ${\cal L}^{(3)} $ and the induced vertices is
essential for making our argument complete. In this way we
show that in the
multi-Regge kinematics the leading contributions of all these
terms result in the multi-Regge effective action with three
relatively simple vertices.

The involved analysis for the other scattering vertex is the
price for the simplicity in obtaining the first two vertices
by working in the particular axial gauge. For extending our
procedure to covariant gauges it would be necessary to
understand, how to introduce the separation of modes in the
presence of redundant fields. The trace of the gauge choice
in our final result is erased, when introducing a complex scalar
field for the scattering gravitons by the non-local relation
$
\phi = - \partial^{-2} h.
$
The multi-Regge effective action for pure gravity can be written
with the Lagrangian \cite{L91}
\begin{eqnarray} 
{\cal L}_{eff} =
- 2 \phi^* (\partial_+ \partial_- -\partial \partial^* )
\partial^2 \partial^{* 2} \phi +
 2 {\cal A}_{++} \partial \partial^* {\cal A}_{--} +
\\  \nonumber
2 \alpha (\partial_- \partial^{* 2} \phi^*
\partial_- \partial^2 \phi ) {\cal A}_{++} \ \ \ +
2 \alpha (\partial_+ \partial^2 \phi^*
\partial_+ \partial^{* 2} \phi ) {\cal A}_{--}
\\     \nonumber
+  \alpha \{ (\partial^{* 2} {\cal A}_{--} \partial^2 {\cal A}_{++}
- \partial \partial^{*} {\cal A}_{--}
\partial \partial^* {\cal A}_{++} ) \phi
+ {\rm c.c.} \}.
\end{eqnarray}
We shall obtain the relation of the fields involved to modes of
the field $h$.

There is no doubt about the gauge independence of the result,
because we know an independent derivation operating merely
with on-shell amplitudes.

\section{Separation of modes}
\setcounter{equation}{0}

The multi-Regge effective action applies to high-energy scattering
$p_A  p_B \rightarrow k_0 k_1 ... k_{n+1} $, where the momenta of
the produced particles obey the conditions of multi-Regge kinematics.
We write these conditions using the notations $s = (p_A + p_B)^2,
s_i = (k_i + k_{i-1} )^2, k_i = q_i - q_{i-1} $ and referring to
the Sudakov decomposition,
\begin{equation}
k^{\mu } = { 1\over \sqrt s } (k_+ p_B^{\mu } + k_-  p_A^{\mu } )
 + \kappa^{\mu }
\end{equation}
as follows
\begin{eqnarray}
s \gg s_i \sim s_j \gg
\vert q_i^2 \vert  \sim \vert q_j^2 \vert,
\ \ \ i,j = 0,1, ..., n+1, \cr
k_{+ i} \ll k_{+  i-1}, \ \ \ k_{- i} \gg k_{- i+1}, \ \ \ \
\prod_{i=1}^{n+1} s_i  =  s \prod_{i=1}^n \vert \kappa_i^2 \vert .
\end{eqnarray}
It is known that the leading logarithmic contribution to the
scattering ampltitudes arises from $s$-channel intermediate states
obeying this condition. Using the effective action we restrict
ourselves to these contributions. Intermediate states not
obeying (2.2) lead to corrections to the effective vertices and
to additional (non-leading) effective vertices \cite{FL}.
We use the complex notation for transverse vectors, e.g. for the
transverse momenta $\kappa = \kappa_1 + i \kappa_2 $, and use
the light-cone components for the longitudinal part of vectors.
The derivatives with respect to coordinates are defined with the
normalization $\partial_+ x_- = \partial_- x_+ = \partial x =
\partial^* x^* = 1$.

We consider the linearized gravity action (1.7).
The gauge vector corresponds to $p_B$.  We separate the
field modes according to the kinematics (2.2),
\begin{equation}
h \rightarrow h_1 + h + h_t,
\end{equation}
where $h_1, h$ and $h_t$ contain correspondingly the modes of the
following momentum ranges
\begin{eqnarray}
h_1 : \ \ \     k_+ k_- \gg \vert \kappa \vert^2
\sim \vert q \vert^2, \cr
h : \ \ \  \vert k_+ k_-  - \vert \kappa \vert^2 \vert
\sim \vert q \vert^2, \cr
h_t : \ \ \ k_+ k_-  \ll \vert \kappa \vert^2
\sim \vert q \vert^2.
\end{eqnarray}
The momenta of $h_t $ are typical for exchanged particles and
the ones of $h$ are typical for scattering particles. In the
generalized leading logarithmic approximation the dominant
contributions correspond to the particles in the $s$-channel
intermediate states strongly ordered in longitudinal momenta and
close to mass shell. Therefore we replace the second line in (2.4)
by
\begin{equation}
h: \ \ \  \vert k_+ k_-  - \vert \kappa \vert^2 \vert
\ll \vert q \vert^2.
\end{equation}
This implies in particular that longitudinal derivatives acting on
$h$ can be approximately replaced by transverse ones:
$ \partial_+ \partial_- h \simeq \partial \partial^* h $.
We keep the notation $h$ for the particular modes of scattering
particles and write $\tilde h$, whenever the other modes are
included.

 The modes $h_1$ corresponding neither to the kinematics
of scattering nor to the one of exchanged particles
are to be integrated out.

We consider first the part of the action (1.7) with kinetic and
the triple interaction terms, ${\cal L}^{(2)} + {\cal L}^{(3)}$.
With the separation (2.3) the kinetic term decomposes,
\begin{equation}
{\cal L}^{(2)} =
- 2 h^*_1 (\partial_+ \partial_- - \partial \partial^* ) h_1
 - 2 h^* (\partial_+ \partial_- - \partial \partial^* ) h
 + 2 h^*_t \partial \partial^*  h_t.
\end{equation}
In the first term the logitudinal part in the d'Alembert operator
clearly dominates. We shall see that (different to the Yang-Mills
case) the contributions proportional to the ratio $ \vert \kappa
\vert^2 / k_+ k_- $ for the modes $h_1$ are important.

In the triple interaction terms the kinematical configuration, where
a field to which the inverse of $\partial_-$ is applied correspond to
a scattering particle with large $k_-$, is suppressed. We denote by
$\tilde h$ the field with all modes and by $\tilde h_t$ fields with modes
$h$ or $h_t$ with momentum components $k_-$ much smaller than the
ones of $\tilde h$ involved in the considered vertex. We introduce
\begin{eqnarray} 
\tilde {\cal A}_{++} = \partial_-^{-2}
(\partial^{* 2} \tilde h_t + \partial^{2} \tilde h_t^* ), \ \ \ \ \
\tilde {\cal A}_{+}^{\prime} = - i \partial_-^{-1}
(\partial^{* 2} \tilde h_t - \partial^{2} \tilde h_t^* ), \cr
\tilde {\cal A}_{+} = 2 \partial_-^{-1} \partial^* \tilde h_t,
\ \ \ \ \ \ \
\tilde {\cal A}_{+}^* = 2 \partial_-^{-1} \partial \tilde h_t^*.
\end{eqnarray}
The tilde will be omitted in the case when only the modes $h_t $
are involved. We define currents as the following bilinear
expressions in $\tilde h$,
\begin{eqnarray} 
\tilde T_{--} = \partial_- \tilde h^* \partial_- \tilde h, \ \ \ \
\tilde T_{-}^* = \frac{1}{2} ( \partial_- \tilde h^*
\partial^* \tilde h +
 \partial^* \tilde h^* \partial_- \tilde h ), \cr
\tilde T_{-} = (\tilde T_-^* )^*,  \ \ \ \ \
\tilde T = \partial \tilde h^* \partial \tilde h, \ \ \ \ \
\tilde T^* = (\tilde T )^*, \cr
J_- = i (\tilde h^* \stackrel{\leftrightarrow}
{\partial}_- \tilde h ),   \ \ \
J^* = i (\tilde h^* \stackrel{\leftrightarrow}
{\partial}^* \tilde h ),  \ \ \
J = (J^*)^*,                     \ \ \ \
j = \tilde h^* \tilde h.
\end{eqnarray}
With these notations the separation of modes $h \rightarrow
\tilde h + \tilde h_t $ leads to the following contribution of
the triple interaction ${\cal L}^{(3)} $,
\begin{eqnarray}
{\cal L}^{(3+)} = 2 \alpha \{
\tilde T_{--} \tilde {\cal A}_{++} - \tilde J_-
\tilde {\cal A}_+^{\prime }
- \tilde T_-^* \tilde {\cal A}_+
- \tilde T_- \tilde {\cal A}_+^*             \cr
+ \tilde T^* \tilde h_t + \tilde T \tilde h_t^*
- i \tilde J^* \partial_- \tilde {\cal A}_+
+ i \tilde J \partial_- \tilde {\cal A}_+^*
- 2 \tilde j \partial_-^2 \tilde {\cal A}_{++}  \}.
\end{eqnarray}
The interaction of a particle with large $k_-$ is dominated by the
first term, giving a contribution ${\cal O}(k_-^2) $. The next three
terms give contributions ${\cal O}(k_-)$ and the remaining ones
contribute to ${\cal O}(k_-^{0})$. There are no helicity flip
vertices up to this accuracy.

The effective vertices for the scattering of a graviton with large
$k_-$ are the contributions of (2.9) when one restricts the currents
to the modes $h$ and the fields ${\cal A}$ to the modes $h_t$.
Here we restrict ourselves to the effective vertices leading to
contributions ${\cal O}(s^2)$ to the amplitudes. The leading
effective vertex for scattering with large $k_-$ is given by the
first term in (2.9) and in the following analysis only the
first four terms will be relevant.

\section{Integration over the heavy modes}
\setcounter{equation}{0}

The essential step in deriving the effective action is the
integration over the modes $h_1$. This will be done approximately
evaluating the action just at the saddle point. To obtain the
main contribution it is enough to consider the action determined by
${\cal L}^{(3+)} + {\cal L}^{(2)}$. The value of the action at the
saddle point is determined by
\begin{equation}
{\cal L}^{(1)} =  2 h_1^{(0) *} (\partial_+  \partial_-  -
\partial \partial^* ) h_1^{(0)},
\end{equation}
where $h_1^{(0)}$ is the solution of the equation obtained by
variation with respect to $h_1^*$,
\begin{eqnarray}
(\partial_+  \partial_- - \partial  \partial^* ) h_1^{(0)} =
\alpha \{ \partial_- (\partial_- h \tilde {\cal A}_{++} )  \cr
+ \frac{1}{2} \partial_-(h \partial^* \tilde {\cal A}_+)
- \frac{1}{2} \partial_-(h \partial \tilde {\cal A}_+^*)
-\partial_- (\partial^* h \tilde {\cal A}_+ )
-\partial (\partial_- h \tilde {\cal A}_+^*)   \}.
\end{eqnarray}
We write the solution inverting formally the d'Alembert operator.
For the modes $h_1$ the term with the longitudinal derivatives is
the leading one, but the next correction has to be kept when
applied to the first term on the r.h.s.  We take into account
that the
momentum component $k_-$ of $\tilde {\cal A}$ is much smaller
than the one of $h$ and that its component $k_+$ is much larger.
Also here we have to keep the first correction proportional to the
ratio of the small to the large $k_+$ components.
\begin{eqnarray}
h_1^{(0)} \simeq \alpha \{
\partial_- h \partial^{-1}_+ \tilde {\cal A}_{++}
- (\partial_+ \partial_- h ) \partial^{-2}_+ \tilde {\cal A}_{++}
+ \partial \partial^* (h \partial^{-2}_+ \tilde {\cal A}_{++}  \cr
+ \frac{1}{2} h \partial_+^{-1} (\partial^* \tilde {\cal A}_+
- \partial \tilde {\cal A}_+^* )
-\partial^* h  \partial_+^{-1} \tilde {\cal A}_+
-\partial h  \partial_+^{-1} \tilde {\cal A}_+^*
- h \partial_+^{-1} \partial \tilde {\cal A}_+^*  \}.
\end{eqnarray}
We see immediately that  in ${\cal L}^{(1)}$
the contribution ${\cal O}(k_-^3)$ in the
product of the leading terms cancels giving up to total derivatives
a result proportional to
\begin{equation}
(\partial_- h^*  \partial_- h) \ \ \partial_- (\partial_+^{-1}
\tilde {\cal A}_{++} \tilde {\cal A}_{++} ).
\end{equation}
This is the reflection of the elementary fact that there is no
dipole radiation in gravity.

Evaluating ${\cal L}^{(1)}$ we keep only terms ${\cal O}(k_-^2)$,
i.e. with two derivatives acting on $h$. We obtain
\begin{eqnarray}
{\cal L}^{(1)} = \alpha^2 T_{--} {\cal I}_{++}, \nonumber    \\
{\cal I}_{++} = - ( \partial_+^{-1} \tilde {\cal A}_{++}
\stackrel{\leftrightarrow}{\partial}_-
\tilde {\cal A}_{++} )  \nonumber \\
+ \{ \partial (
\partial_+^{-1} \tilde {\cal A}_{++} \tilde {\cal A}_+^*
- \tilde {\cal A}_{++}  \partial_+^{-1} \tilde {\cal A}_+^* )
+ \partial \tilde {\cal A}_{++} \partial_+^{-2}
\partial^* \tilde {\cal A}_{++}
+ {\rm c.c. } \}.
\end{eqnarray}
We have expected the result to be proportional to the current
$T_{--}$. In the calculation it emerges from the cancellation
of many terms
with other structures.

Now we interprete (3.5) as the quartic terms emerging from the
integration over the $t$-channel modes $h_t$ with an action
determined by the kinetic term, the leading effective scattering
vertex from (2.9) and the induced triple vertex
\begin{equation}
{\cal L}^{(1)}_{ind}
= - \frac{1}{2} \alpha
\partial \partial^* {\cal A}_{--} {\cal I}_{++},
\end{equation}
where we introduced the notation
\begin{eqnarray}
{\cal A}_{--} = \partial_-^2 (\partial \partial* )^{-2}
(\partial^{* 2} h_t + \partial^2 h_t^* ) =
\partial_-^4 (\partial \partial* )^{-2} {\cal A}_{++}, \cr
{\cal A}_{-}^{\prime } = - i \partial_- (\partial \partial* )^{-2}
(\partial^{* 2} h_t - \partial^2 h_t^* ).
\end{eqnarray}

\section{The effective production vertex}
\setcounter{equation}{0}

The contribution of the triple interaction terms to the
configuration where one field carries a component $k_- $ much
smaller than the $k_-$ of the other two fields (2.9) involves
two cases. One is the case of scattering, where both fields in
the currents are of type $h$. The other is the case of production,
where one field in the currents is of type $h_t$ and the other
of type $h$.

We write the contribution of the leading term in (2.9)
in the second case using the notations (3.7),
\begin{equation}
{\cal L}^{(3-+)} =
- \alpha \{ \partial^{* 2 } ({\cal A}_{--} - i \partial_-
{\cal A}_-^{\prime } ) h + {\rm c.c.} \} {\cal A}_{++}.
\end{equation}
The contribution with ${\cal A}_-^{\prime }$ is irrelevant for
the leading effective vertices.

Also the induced vertex (3.6) contributes to production.
In this case one of the fields in ${\cal I}_{++}$ carries the
modes $h$ (${\cal A}^{(s)}$) and the other the modes $h_t$ ($
{\cal A}$). The contributions where the modes $h_t$ are in
$\partial_- {\cal A}_{++}, \partial_+^{-1} {\cal A}_{++}$ or
in ${\cal A}_+, {\cal A}_+^* $ are small. Thus we have
\begin{eqnarray}
{\cal L}^{(1-+)}_{ind} = - \alpha \partial \partial^*
{\cal A}_{--}
[ \partial_- \partial_+^{-1} {\cal A}^{(s)} {\cal A}_{++}
\nonumber  \\
+ \{ - \partial (\partial_-^{-1} {\cal A}_+^{(s) *} {\cal A}_{++} )
+ \partial_+^{-2} \partial^* {\cal A}_{++}^{(s)}
\partial {\cal A}_{++} + {\rm c.c. } \} ].
\end{eqnarray}
Using (2.5) and (2.7) we rewrite this as
\begin{equation}
{\cal L}^{(1-+)}_{ind} =  \alpha \partial \partial^*
{\cal A}_{--}
[ \partial^* \partial^{-2} h  \partial {\cal A}_{++}
- \partial^* ( \partial^{-1} h {\cal A}_{++} ) + {\rm c.c.} ].
\end{equation}
We obtain the effective production vertex as the sum of
(4.1) and (4.3).
\begin{equation}
{\cal L}^{(-+)} = - \alpha (
\partial^{* 2} {\cal A}_{--} \partial^2 {\cal A}_{++} -
\partial \partial^{*} {\cal A}_{--}
\partial \partial^* {\cal A}_{++}  )
\partial^{-2} h + {\rm c.c.}
\end{equation}

The quartic terms do not give a leading contribution to the
production. The elastic and also the inelastic scattering
of a particle with large $k_-$ in
an external gravitational field is determined only by the triple
interaction terms.

\section{The quartic interaction terms}
\setcounter{equation}{0}

To lowest order the peripheral scattering of two quanta
at high energy is
determined by the quartic and the triple interaction terms. The
latter contribution is obtained by contracting one vertex in the
kinematics (2.2) with another one, where the field of type $h_t$
now has a momentum component $k_-$ of the same order as the
largest $k_-$ of the other two fields involved. Let us extract
from ${\cal L}^{(3)}$ (1.7) the contribution to this
kinematical configuration,
${\cal L}^{(3-)} = {\cal L}^{(3- 1)} + {\cal L}^{(3- 2)}$.
Each of the three fields in the vertex can be in the modes $h_t$.
${\cal L}^{(3- 1)}$ corresponds to the case where the field with
the inverse derivative in ${\cal L}^{(3)}$ (1.7) is $h_t$. This
contribution may seem unnatural. Indeed we shall see now that it
just cancels to a large extend the contribution from the quartic
terms ${\cal L}^{(4)}$ to particle-particle scattering.
\begin{eqnarray}
{\cal L}^{(3- 1)} = 2 \alpha \{
T_{--} {\cal A}_{++} - J_- {\cal A}_+^{\prime }
- T_-^* {\cal A}_+   - T_- {\cal A}_+^*  \cr
- \frac{i}{2} J^* \partial {\cal A}_+
  + \frac{i}{2} J \partial {\cal A}_+^*
- \frac{1}{2} j \partial_-^2 {\cal A}_{++}  \}.
\end{eqnarray}

We consider an action determined by ${\cal L}^{(3+)}$ (2.9),
${\cal L}^{(3- 1)}$ (5.1) and the kinetic term and integrate
over $h_t$. We keep only the terms arising from the leading
term in ${\cal L}^{(3+)} $, because we are interested in the
${\cal O}(k_-^2)$ contributions to the scattering only. In the
kinematics for which ${\cal L}^{(3+)}$ is written the momenta
$k_-$ of the fields in ${\cal L}^{(3-)}$ are much smaller. We
obtain
\begin{eqnarray}
{\cal L}^{(3- 1)}_{ind} = - 4 \alpha^2 T_{--} \{
\partial \partial^* \partial_-^{-4} T_{--}
-  \partial_-^{-3}  ( \partial T_{-}^* +
 \partial^* T_{-} ) \cr
- \frac{1}{2} \partial_-^{-2} \partial \partial^* j
+ \frac{i}{4} \partial_-^{-2} ( \partial J^* - \partial^* J ) \}
+ {\cal O}(k_-) .
\end{eqnarray}

 The contribution of the quartic terms to scattering becomes more
transparent when we rewrite them using the currents (2.8) as
\begin{eqnarray} 
{\cal L}^{(4)} = 2 \alpha^2 \{
+ \vert \partial_-^{-2} [ \partial^* T_{--} -
 \partial_- T_-^* -
\frac{i}{4} \partial_- \partial^* J_-
+ \frac{i}{4} \partial_- ^2 J^* ] \vert^2
\cr
- 2 \vert \partial_-^{-2} [ \partial^* T_{--}
- \partial_- T_-^*
- \frac{3 i}{4} \partial_- \partial^* J_-
+ \frac{3 i}{4} \partial_-^2 J^*
- \partial_-^2 \partial^* j
-\partial^3 (h^* \partial^* \partial_-^{-1} h ) ] \vert^2
\cr
+ \vert \partial_-^{-1} [
\frac{i}{2} \partial^* J_- \frac{i}{2} \partial_- J^* ] \vert^2
- 3 \vert \partial_-^{-1} [
 T_-^* + \frac{i}{4} \partial^* J_-
 - \frac{i}{4} \partial_- J^* ] \vert^2
\cr
-3 \partial_-^{-2} T_{--} (\partial h^* \partial^* h ) +
(\partial_-^{-2} T_{--} - j)
[ \partial \partial^* j -
\partial_- ( \partial \partial^* \partial_-^{-1} h^*  h +
h^* \partial \partial^* \partial_-^{-1} h ) ] \}.
\end{eqnarray}
 We pick up the terms giving a contribution ${\cal O}(k_-^2)$ to the
scattering,
\begin{eqnarray} 
{\cal L}^{(4)} = 2 \alpha^2  T_{--} \{
 \partial \partial^* \partial_-^{-4} T_{--}
- \frac{1}{2} \partial_-^{-3} (\partial T_-^* + \partial^* T_- )
\cr
- 3 \partial_-^{-2} \partial \partial^* j +
\frac{5 i}{4} \partial_-^{-2} ( \partial J^* -\partial^* J )
-3 \partial^{-2} (\partial h^* \partial h)
\cr
+  [ \partial^* \partial_-^{-1} ( \partial \partial_-^{-1} h^* h )
+ \partial \partial_-^{-1} (h^* \partial^* \partial_-^{-1} h ) -
\partial \partial_-^{-1} h^*  \partial^* \partial_-^{-1} h ]
\cr
- \frac{1}{2} \partial_-^{-1}
( \partial \partial^* \partial_-^{-1} h^* h
 + h^* \partial \partial^* \partial_-^{-1} h ) \}.
\end{eqnarray}

In (5.2) we wrote only the contributions, where $T_{--}$ carries the
large $k_-$. Here in (5.3) each of the two currents in every
term can carry the large momentum $k_-$. Therefore the first
term in (5.4) contributes twice and its contribution thus
cancels the one of the first term in (5.2) in the sum.
The contributions from (5.3) where a non-leading current carries
the large $k_-$ can be disregarded.

We analyse ${\cal L}^{(4)} + {\cal L}^{(3- 1)}_{ind} $ using the
fact that the fields in the curly brackets
in (5.2) and (5.4) carry relatively
small $k_-$ and large $k_+$ and that the modes $h$ obey (2.5).
First we observe that the square bracket in (5.4) is approximately
equal to $\partial_-^{-1} \partial_+ j$ and therefore does not
give a contribution ${\cal O}(s^2)$ to the scattering. Also the
last term in (5.4) is approximately proportional to the latter
expression. Further we have in the considered kinematics
\begin{equation}
\partial T_{-}^* + \partial^* T_{-} = \partial_+ T_{--} +
\frac{1}{2} \partial_- ( \partial^* h^*  \ \partial h +
 \partial h^*  \ \partial^* h ),
\end{equation}
where again the first term is neglegible. This allows to write
the leading contributions as
\begin{equation}
{\cal L}^{(4)} + {\cal L}^{(3- 1)}_{ind} = - 2 \alpha^2
T_{--} \partial_-^{-2}
(\partial \partial^* h^* h + h^* \partial \partial^* h +
3 \partial h^* \partial^* h ).
\end{equation}

We interprete the result as arising from intergrating over $h_t$
with an action determined by the leading term of ${\cal L}^{(3+)}$
(2.9), the kinetic term and the induced triple vertex
\begin{equation}
{\cal L}^{(3-4)}_{ind} =  2 \alpha
\partial \partial^* \partial_-^{-2} {\cal A}_{--}
(\partial \partial^* h^* h + h^* \partial \partial^* h +
3 \partial h^* \partial^* h ).
\end{equation}

Including this induced vertex we have to remove the contribution
${\cal L}^{(3- 1)}$, i.e the one where the fields with
$\partial_-^{-1}$ play the role of ${\cal A}_{--}$,
 and the quartic terms ${\cal L}^{(4)}$ from which
it was generated. Therefore we have now two types of exchanged
fields; writing ${\cal A}_{++}$ and ${\cal A}_{--}$ are not any
more convenient notations for the same object. In the graphs the
exchange lines obtain arrows related to the longitudinal
momentum ordering.
When ${\cal A}_{++}$ and ${\cal A}{--} $ become independent the
normalization of their kinetic term changes by a factor of 2.

\section{The other effective scattering vertex}
\setcounter{equation}{0}

We write now the second contribution ${\cal L}^{(3- 2)}$ of the
original triple vertex to the configuration involving one field
$h_t$ carrying relatively large $k_-$,
\begin{eqnarray} 
{\cal L}^{(3- 2)} =  \alpha \{
\partial^{* 2} ({\cal A}_{--} - i \partial_- {\cal A}_{-}^{\prime} )
[ - h \partial^{* 2} \partial_-^{-2} h ] \nonumber  \\
- 2 \partial_-^{-1} \partial^{*} ({\cal A}_{--} -
i \partial_- {\cal A}_{-}^{\prime} )
[\partial^*  h \partial^{*} \partial_-^{-1} h ] \nonumber  \\
- \partial_-^{-2} \partial^{2} ({\cal A}_{--} +
 i \partial_- {\cal A}_{-}^{\prime} )
[\partial_-^2 h^* \partial^{* 2} \partial_-^{-2} h ] \nonumber \\
-\partial_-^{-2} \partial^* \partial^{2} ({\cal A}_{--} +
 i \partial_- {\cal A}_{-}^{\prime} )
[ \partial_- h^* \partial^{*} \partial_- h ]    + {\rm c.c.} \}
\end{eqnarray}
The contributions with ${\cal A}_-^{\prime }$ can be ignored here.

We write also the contribution ${\cal L}_{ind}^{(-)}$
of the induced vertex
${\cal L}^{(1)}_{ind}$  (3.6) to the scattering, i.e. to the case
where both fields in ${\cal I}_{++}$ carry modes $h$ with
$k_+$ relatively large. We apply (2.5) and (2.7) and disregard
contributions which do not give the second power of the large
$k_+$.
\begin{eqnarray}
{\cal L}^{(-)}_{ind} = - \frac{1}{2} \alpha
\partial \partial^* {\cal A}_{--}
\{ [ \partial_+^2 ( \partial^{-1} h + \partial \partial^{* -2}
h^* )
(\partial^* \partial^{-2} h - 3 \partial^{* -1} h^* )  +
{\rm c.c.} ] \cr
- 2 \partial_+^2 ( \partial^{-2} h + \partial^{* -2} h^* )
(\partial^* \partial^{-1} h + \partial \partial^{* -1} h^* )
\}.
\end{eqnarray}

We extract first the terms, which would contribute to
helicity-flip scattering and show that they cancel in the sum.
We apply (2.5) and obtain from (6.1)
\begin{eqnarray}
{\cal L}^{(3-2)} \vert_{hh} =
 \alpha \partial^{* 2}
{\cal A}_{--}
[ - h \partial^{* 2} \partial_-^{-2} h +
\partial^* \partial_-^{-1} h \ \partial^* \partial^{-1} h ]  \cr
= -  \alpha \partial^{* 2} {\cal A}_{--}
\partial [ \partial_+^2 \partial^{-2} h \partial^{-1} h ].
\end{eqnarray}
Disregarding a term proportional to $\partial_+ {\cal A}_{--}$
we have from (6.2)
\begin{equation}
{\cal L}^{(-)}_{ind} \vert_{hh} =  \alpha \partial \partial^*
{\cal A}_{--}
[ \partial^* \partial_+^2 \partial^{-2} h \partial^{-1} h +
  \partial_+^2 \partial^{-2} h \partial^* \partial^{-1} h ],
\end{equation}
which indeed cancels in the sum against (6.3).

Now we look at the helicity conserving terms in
${\cal L}^{(3- 2)}$  (6.1),
\begin{eqnarray}
{\cal L}^{(3- 2)} \vert_{h^*h} = -  \alpha {\cal A}_{--}
[ \partial^2 (h^* \partial^{* 2} \partial_-^{-2} h ) +
{\rm c.c.} ]   \cr
- 2 \partial_-^{-1} {\cal A}_{--}
[ \partial^* \partial^2 ( h^* \partial^* \partial_-^{-1} h )
 \partial^2 ( h^* \partial^{* 2} \partial_-^{-1} h )
+ {\rm c.c.}  ] \cr
- 2 \partial_-^{-2} {\cal A}_{--}
[ 3 \partial^2 \partial^* (h^* \partial^* h ) -
\partial^2 ( \partial^* h^* \partial^* h )
+ {\rm c.c.} ].
\end{eqnarray}
Using (2.5) and disregarding total $\partial_+$ derivatives we
transform the second quare bracket as
\begin{eqnarray}
\partial^* ( \partial^2 h^* \partial^* \partial_-^{-1} h +
\partial h^* \partial \partial^* \partial_-^{-1} h +
h^* \partial^* \partial^2 \partial_-^{-1} h +
\partial^* h \partial^2 \partial_-^{-1} h^* ) + {\rm c.c.} \cr
+ \partial_- [ \partial^* ( \partial^2 \partial_-^{-1} h ) +
{\rm c.c.} ] + \partial_+ (...).
\end{eqnarray}
In the same way we obtain the relation
\begin{eqnarray}
\partial \partial^* (h^* h) + {\rm c.c.} =
2 ( \partial \partial^* h^* \partial \partial^* h ) -
\partial_-^{2} (\partial_+ h^* \partial_+ h) +
\partial_+ (...),
\end{eqnarray}
which can be used to transform the second term in the last
bracket of (6.5) as
\begin{eqnarray}
\partial^2 (\partial^* h^* \partial^* h ) + {\rm c.c.} = \cr
2 \partial \partial^* h^* \partial \partial^* h +
[ \partial \partial^* (\partial \partial^* h^* h ) -
(\partial \partial^* )^2 h^* h + {\rm c.c.} ] =  \cr
- \partial_-^{2} ( \partial_+ h^* \partial_+ h ) +
 \partial \partial^* ( \partial \partial^* h^* h +
{\rm c.c.} ).
\end{eqnarray}
Therefore (6.5) reduces to
\begin{eqnarray}
{\cal L}^{(3- 2)} \vert_{h^* h} =  \alpha \{
{\cal A}_{--} [ - \partial^2 ( h^* \partial^{* 2}
\partial_-^{-2} h )                          \cr
+ \frac{1}{2} (\partial_+ h^* \partial_+ h ) +
2 \partial^* ( \partial^2 \partial_-^{-1} h^*
\partial^* \partial_-^{-1} h ) + {\rm c.c.} ]   \cr
- \partial_-^{-2} \partial \partial^* {\cal A}_{--}
[ 3 \partial ( h^* \partial^* h ) -
( \partial \partial^* h^* h) + {\rm c.c.} ] \}.
\end{eqnarray}
The second term cancels in the sum with the induced vertex
${\cal L}^{(3-4)}_{ind} $ (5.7)  resulting from the cancellation
between the quartic terms and the contribution of
${\cal L}^{(3- 1)}$. The first term in (6.9) has to be added to
the helicity conserving
contribution of ${\cal L}^{(-)}_{ind}$ (6.2) from the
induced vertex
arising from the integration over $h_1$. The latter can be written
as
\begin{eqnarray}
{\cal L}^{(-)}_{ind} \vert_{h^* h} = - \alpha {\cal A}_{--} \{
\frac{1}{2} (\partial \partial^* )^2
[ \partial_+^2 \partial^{* -2} h^* \partial^{-2} h ] \cr
- 2 \partial^2 \partial^* [ \partial_+^2 \partial^{* -1} h^*
\partial^{-2} h ]  +{\rm c.c.} \}
\cr
 = - \alpha {\cal A}_{--} \{
\partial^2 \partial_+ h^* \partial_+ \partial^{-2} h
+ 4 \partial_+ \partial h^* \partial_+ \partial^{-1} h  \cr
+ 2 \partial_+ \partial \partial^{* -1} h^*
\partial_+ \partial^* \partial^{-1} h
+ \frac{3}{2} \partial_+ h^* \partial_+ h \cr
- \partial_+ \partial^2 \partial^{* -2} h^*
 \partial_+ \partial^{* 2} \partial^{-2} h   + {\rm c.c.} \}
\end{eqnarray}
We transform the first term in (6.9) using (2.5),
\begin{eqnarray}
{\cal L}^{(3- 2)} \vert_{h^* h} + {\cal L}^{(3-4)}_{ind} =
 \alpha {\cal A}_{--} \{
- \partial^2 [ h^* \partial_+^2 \partial^{-2} h ]  \cr
+ 2 \partial^* (\partial \partial^{* -1} \partial_+ h^*
\partial_+ \partial^{-1} h )
+ \frac{1}{2} (\partial_+ h^* \partial_+ h )
+ {\rm c.c.} \}
\end{eqnarray}
This cancels the contribution of ${\cal L}^{(-)}_{ind}\vert_{h^* h} $
up to the last term in (6.10). This remaining
term is the effective vertex
for scattering of partices with large $k_+$ in the gauge (1.2).
Transforming to the gauge, where the other incoming particle
momentum $p_A$ plays the role of the gauge vector (i.e. exchanging
indices $+$ and $-$ in (1.2) ), the result looks similar to the
first effective scattering vertex with the indices $+$ and $-$
exchanged.
\begin{eqnarray}
{\cal L}^{(3- 2)} + {\cal L}^{(3-4)}_{ind} +
  {\cal L}^{(-)}_{ind}  &=
2 \alpha {\cal A}_{--}  (
 \partial_+ \partial^2 \partial^{* -2} h^*
 \partial_+ \partial^{* 2} \partial^{-2} h ) \cr &=
 2 \alpha {\cal A}_{--} (\partial_+ h^{(g) *}
\partial_+ h^{(g)} ).
\end{eqnarray}
With the results (2.9), (4.4) and (6.12) for the effective
vertices we recover
the known result (1.8) for the effective action.

\section{Conclusions}

We have established the direct relation of the multi-Regge
effctive action in gravity to the Einstein - Hilbert action.
The method of separating modes according to the multi-Regge
kinematics and of integrating over "heavy" modes worked out
first for the Yang-Mills case has been extended
to the case of gravity.
In an axial gauge with the redundant metric components
excluded we established the relations of the fields describing
scattering ($\phi , \phi^* $) and exchanged (${\cal A}_{++},
{\cal A}_{--}$ ) gravitons to momentum modes of the two
independent physical fields of the metric fluctuation
($h, h^* $).  The fields describing exchanged gravitons
can be considered as pre-reggeons.

The extension turned out to be more than a straightforward
exercise. We had to understand the contributions of involved
interaction terms to the peripheral high-energy scattering.
Compared to the Yang-Mills case higher accuracy is required
in the ki\-ne\-ma\-ti\-cal approximations referring to the
 momentum
orderings imposed by the multi-Regge kinematics and in
particular by the mode separation. The physical reason for
this is simply the absence of dipole radiation.

Being an effective action the applicability of (1.8) is
restricted clearly to the ki\-ne\-ma\-ti\-cal
region for which it
has been derived. For example, the exchanged fields
${\cal A}$ have a purely transverse  propagator. But this
would generate wrong contributions, if one forgets about
the condition (2.4) on the momentum range. Actually these
conditions should be incorporated into the action by
damping factors or by cut-offs.

In the case of QCD we have understood earlier \cite{KLS} how
to deal with fermions in the derivation of the effective action.
Combining this with the experience from the present analysis
it will be not difficult to generalize it to
supergravity and to the coupling of gravity to matter.

We have restricted ourselves to
contributions resulting in the leading effective vertices
which are related to the exchanges of gravitons contributing
each with an addition power of $s$. Including $s$-channel
intermediate gravitons results in corrections proportional to
one power of $\ln s $ for each loop. In view of this it would
be desirable to include into the effective action also the
non-leading graviton exchanges, which do not change the power of
$s$ of a given contribution to the amplitude. The corresponding
effective scattering vertices are the second to fourth terms
in ${\cal L}^{(3+)}$ (2.9). However this requires a further
improvement of the kinematical approximations.

There are ideas about the extension of the approach to
covariant gauges, which should be tried first in the
Yang-Mills case. Also the approximation used here in the
integration over the "heavy" modes can be improved with
some effort.

It is important to extend the analysis to the full gravity
action, including the terms of all orders in the
metric fluctuation $h$. We see now good reasons to hope
that this can be done.

\vspace*{3cm}
\newpage

$\mbox{ }$ \\
{\Large\bf Acknowledgements} \\
$\mbox{ }$ \\
We are grateful to D. Amati and L.N. Lipatov for useful
discussions.

We thank the Volkswagen Stiftung for supporting our collaboration
in a difficult si\-tu\-ation.  One author (R.K.) is grateful
to SISSA for kind hospitality and the other (L.Sz.) acknowledges
the support by Komitet Badan Naukowych 2P302 143 06.

\end{document}